\documentclass[12pt]{iopart}

\usepackage{amssymb}
\usepackage{iopams}
\usepackage{graphicx}
\usepackage[numbers,sort]{natbib}

\graphicspath{{pict/}{}}

\newcommand\pictc[5]{\begin{figure}
                   \centerline{
                   \includegraphics[width=#1\columnwidth,height=0.8\textheight,keepaspectratio]{#3}}
               \protect\caption{\protect\label{fig:#4} #5}
                \end{figure}            }

\newcommand\pict[4][0.75]{\pictc{#1}{!tb}{#2}{#3}{#4}}
\newcommand\pictb[4][0.75]{\pictc{#1}{!b}{#2}{#3}{#4}}

\newcommand\rpict[1]{\ref{fig:#1}}

\newcommand\leqt[1]{\protect\label{eq:#1}}
\newcommand\reqtn[1]{\ref{eq:#1}}
\newcommand\reqt[1]{(\reqtn{#1})}

\newcommand\lsect[1]{\protect\label{sect:#1}}
\newcommand\rsect[1]{\ref{sect:#1}}

\begin{document}

\title{Effect of loss on photon-pair generation in nonlinear waveguides arrays}

\author{Diana A. Antonosyan, Alexander S. Solntsev and Andrey~A.~Sukhorukov}

\address{Nonlinear Physics Centre, Research School of Physics and Engineering, Australian National University, Canberra 0200 Australia}
\ead{diana.antonosyan@anu.edu.au}
\begin{abstract}
We describe theoretically the process of spontaneous parametric down-conversion in quadratic nonlinear waveguide arrays in the presence of linear loss.
We derive a set of discrete Schrodinger-type equations for the biphoton wave function, and the wave function of one photon when the other photon in a pair is lost. We demonstrate effects arising from loss-affected interference between the generated photon pairs and show that nonlinear waveguide arrays can serve as a robust loss-tolerant integrated platform for the generation of entangled photon states with non-classical spatial correlations.
\end{abstract}


\maketitle

\section{Introduction}

Optical quantum communications and computation schemes rely on controlled preparation of well-defined photonic states~\cite{Bouwmeester:2000:QuantumInformation, Shih:2011:IntroductionQuantum}. Spontaneous parametric down-conversion (SPDC) in nonlinear crystals~\cite{Harris:1967-732:PRL, Klyshko:1988:PhotonsNonlinear, Mandel:1995:OpticalCoherence} has become a source of choice for experimental generation of correlated and entangled photon pairs with demonstrations of such effects as quantum teleportation~\cite{Bouwmeester:1997-575:NAT, Boschi:1998-1121:PRL,
Pan:1998-3891:PRL, Furusawa:1998-706:SCI}, quantum cryptography~\cite{Ekert:1992-1293:PRL}, Bell-inequality violations~\cite{Giustina:2013-227:NAT} and quantum imaging~\cite{Fickler:2013-1914:SRP}.

The mode confinement in a waveguide enables a significant increase of the SPDC source brightness in comparison to bulk crystal setups~\cite{Tanzilli:2001-26:ELL}. Even more importantly, waveguide integration provides interferometric stability, which is essential for quantum simulations and cryptography. SPDC in nonlinear waveguides can be implemented to produce photon pairs in distinct spatial modes~\cite{Banaszek:2001-1367:OL, Zhang:2007-10288:OE, Zhang:2010-64401:JJAP, Eckstein:2008-1825:OL, Saleh:2010-736:IPJ}. Overall, nonlinear waveguides can serve as photon-pair sources ideally suited for applications in quantum communications~\cite{URen:2003-480:RAR}.

Recently, there has been growing interest in the study of the propagation of nonclassical light in coupled waveguides:  quantum gates were implemented using pairs of waveguides acting as integrated beam splitters~\cite{Politi:2008-646:SCI}, and lattices of coupled waveguides were used for the study of Bloch oscillations~\cite{Longhi:2008-193902:PRL} and propagation of squeezed light~\cite{Rai:2008-42304:PRA}.
Overall integrated optical quantum circuits utilising coupled waveguides are increasingly gaining attention as a possible solution for scalable quantum technologies with important applications to quantum simulations. A key mechanism for quantum simulations can be provided by the process of quantum walks in an optical waveguide array (WGA)~\cite{Peruzzo:2010-1500:SCI}, with applications to boson sampling~\cite{Broome:2013-794:SCI, Spring:2013-798:SCI, Tillmann:2013-540:NPHOT, Crespi:2013-545:NPHOT}. Furthermore, it was recently suggested~\cite{Solntsev:2012-23601:PRL, Solntsev:2012-27441:OE, Grafe:2012-562:SRP, Kruse:2013-83046:NJP} that a nonlinear waveguide array can be used for both photon-pair generation through spontaneous parametric down-conversion and quantum walks of the generated biphotons with strong spatial entanglement between the waveguides. Importantly, such integrated scheme avoids input losses, since in an integrated nonlinear waveguide array photon pairs can be generated inside the quantum walk circuit. The internal losses in the waveguides however may still be present. In this work, we address an important question of the tolerance of the biphoton generation to possible losses in the waveguides.
We focus our attention on Markovian losses, as this is the most common type of losses in waveguides, which can be associated in particular with leaky modes~\cite{Ghatak:1985-311:OQE, Zhu:2006-1619:JLT, Rai:2012-52330:PRA}.

This paper is organised as follows. Section~\rsect{ModelS} contains detailed investigation of spontaneous parametric down-conversion in a single lossy quadratic nonlinear waveguide. We explore the dependence of photon-pair intensity on losses and  phase mismatch and demonstrate a number of counter-intuitive effects. For example we show that the increase in idler losses can lead to the increase of signal intensity, and that the signal intensity becomes independent on nonlinear waveguide length after a particular propagation distance. We also demonstrate that signal and idler losses lead to the transformation of common ${\rm sinc}$-shaped photon-pair correlation spectrum into a Lorenzian shape, and that this transformation can be fully reversed by the specific increase in pump losses. The results related to WGAs are presented in Sec.~\rsect{WGAS}. We derive a model of the SPDC and photon-pair propagation in finite quadratic nonlinear WGAs with losses and present the detailed analysis of the generated photon-pair spatial correlations, entanglement and spatial intensity distributions. We show that photon-pair spatial entanglement generated in nonlinear WGAs remains strong even in the presence of high losses.

\section{Spontaneous parametric down-conversion in a single $\chi^{\left(2\right)}$ waveguide with losses}\lsect{ModelS}

The process of SPDC can occur in a $\chi^{\left(2\right)}$ nonlinear waveguide pumped by a pump laser, where a pump photon at frequency $\omega_{p}$ can be spontaneously split into signal and idler photons with corresponding frequencies $\omega_{s}$ and $\omega_{i}$, such that $\omega_{p}=\omega_{s}+\omega_{i}$.
The effect of linear losses on SPDC was previously considered in various contexts~\cite{Caves:1987-1535:JOSB, Klyshko:1988:PhotonsNonlinear, Kulik:2004-31:JETP}. Here, we perform a detailed analysis of the emerging photon intensities and correlations, in the regime of photon-pair generation.

To describe waveguide losses, it is possible to introduce them through series of virtual asymmetric beam-splitters in an otherwise conservative medium~\cite{Caves:1987-1535:JOSB, Loudon:2000:QuantumTheory}, see Fig.~\rpict{1WGM}. At each step during propagation from $z$ to $z+\Delta z$ the photon pairs can be generated through SPDC. On the other hand there is a probability for signal and idler photons to be reflected by beam-splitters, corresponding to the loss of photons from the waveguide. Then, according to the general principles~\cite{Lloyd:1996-1073:SCI}, the photon dynamics is governed by a sum of Hamiltonians which individually describe SPDC in lossless nonlinear medium ($\hat{H}_{nl}$) and linear losses due to virtual beam-splitters ($\hat{H}_{bs}$), $\hat{H} = \hat{H}_{nl} + \hat{H}_{bs}$.

\pict{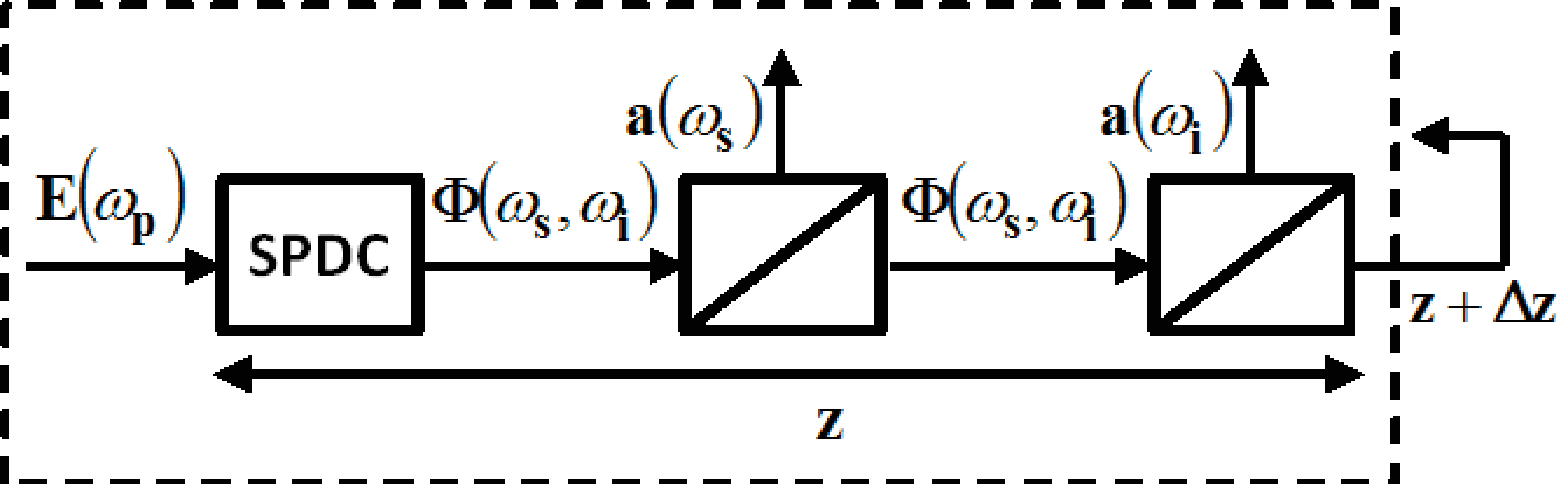}{1WGM}{Scheme of photon-pair propagation involving SPDC and losses in a single waveguide.
The first step shows the probability of photon-pair generation though SPDC, the second step corresponds to the probability to lose a signal photon, and the third step corresponds to the probability to lose an idler photon.}

The SPDC process in the absence of losses, in the undepleted classical pump approximation, is governed by a Hamiltonian~\cite{Mandel:1995:OpticalCoherence}:

\begin{eqnarray} \leqt{Hnl}
  &&\hat{H}_{nl}(z) = \int d \omega_s \beta_s^{(0)}(\omega_s) a_s^\dag(\omega_s) a_s(\omega_s)
            +  \int d \omega_i \beta_i^{(0)}(\omega_i) a_i^\dag(\omega_i) a_i(\omega_i) \\
        &&+ \int d \omega_s \int d \omega_i  \left[ E_p(z, \omega_s+\omega_i) a_s^\dag(\omega_s) a_i^\dag(\omega_i)
                                                   + E_p^\ast(z, \omega_s+\omega_i) a_s(\omega_s) a_i(\omega_i) \right], \nonumber
\end{eqnarray}
where $a_{s,i}^\dag$ and $a_{s,i}$ are the creation and annihilation operators for the signal and idler photons with the commutators $[a_s(\omega_1), a_s^\ast(\omega_2)]= \delta(\omega_1-\omega_2)$ and $[a_i(\omega_1), a_i^\ast(\omega_2)]= \delta(\omega_1-\omega_2)$, $\delta(z)$ is a Kronecker delta-function, $E_p(z, \omega_p)$ is proportional to the pump amplitude at frequency $\omega_p$ and quadratic nonlinearity, and $\beta_{s,i}^{(0)}$ are the signal and idler propagation constants relative to the pump.

We assume Markovian losses and negligible thermal fluctuations.
Then, the Hamiltonian corresponding to a series of beam-splitters~\cite{Prasad:1987-139:OC, Caves:1987-1535:JOSB, Loudon:2000:QuantumTheory} can we written as:
\begin{eqnarray} \leqt{Hbs}
  \hat{H}_{bs}(z) &=& \int d \omega_s \sqrt{2 \gamma_s(\omega_s)} \left[ a_s(\omega_s) b_s^\dag(z, \omega_s)
                                                + a_s^\dag(\omega_s) b_s(z, \omega_s) \right]  \\
                      &+&   \int d \omega_i \sqrt{2 \gamma_i(\omega_i)} \left[ a_i(\omega_i) b_i^\dag(z, \omega_i)
                                    + a_i^\dag(\omega_i) b_i(z, \omega_i) \right],  \nonumber
\end{eqnarray}
where the operators $b_{s,i}^\dag(z, \omega)$ describe creation of photons which are lost from a waveguide after a beam-splitter at coordinate $z$, with the commutators $[b_s(z_1, \omega_1), b_s^\ast(z_2, \omega_2)] = \delta(z_1-z_2) \delta(\omega_1-\omega_2)$ and $[b_i(z_1, \omega_1), b_i^\ast(z_2, \omega_2)]= \delta(z_1-z_2) \delta(\omega_1-\omega_2)$, and $\gamma_{s,i}$ are the linear loss coefficients.

We focus on the generation of a single photon pair and consider multi-photon-pair processes to be negligible for appropriately attenuated pump power. Then, the generation of photon pairs with different frequencies occurs independently, due to the absence of cascading processes. We will therefore omit $\omega_{s,i,p}$ in the following analysis to simplify the notations.
Then, we seek a solution for a two-photon state at distance $z$ as:
\begin{eqnarray}
        |\Psi(z)\rangle &=& \Phi(z) a_s^\dag a_i^\dag |0\rangle
         + \int_0^z d z_l \tilde{\Phi}^{(s)}(z, z_l) a_s^\dag b_i^\dag(z_l)  |0\rangle \nonumber\\
        & + & \int_0^z d z_l \tilde{\Phi}^{(i)}(z, z_l) b_s^\dag(z_l) a_i^\dag  |0\rangle \leqt{state1wg} \\
        & + & \int_0^z d z_{l_s} \int_0^z d z_{l_i} \tilde{\Phi}^{(si)}(z_{l_s}, z_{l_i}) b_s^\dag(z_{l_s}) b_i^\dag(z_{l_i})  |0\rangle , \nonumber
\end{eqnarray}
where $|0\rangle$ denotes a vacuum state with zero number of signal and idler photons.
The equation for the evolution of the state vector is  ${d \Psi(z)}/{d z} = - i \hat{H}(z) [|0\rangle + |\Psi(z)\rangle]$, assuming undepleted vacuum state. Then, we obtain the following equations for the two-photon wave functions:
\begin{eqnarray}
    \frac{\partial \Phi(z)}{\partial z}= - (i \Delta\beta^{(0)} + \gamma_{s}+\gamma_{i})\Phi(z)
                                +A \e^{-\gamma_{p}z}, \quad \Phi(z=0) = 0, \leqt{dPhi} \\
    \frac{\partial\tilde{\Phi}^{(s)}(z,z_l)}{\partial z} = -(i \beta_s^{(0)} + \gamma_{s}) \tilde{\Phi}^{(s)}(z,z_l)=0,\, z \ge z_l , \leqt{dPhi_s}\\
    \frac{\partial\tilde{\Phi}^{(i)}(z,z_l)}{\partial z} = -(i \beta_i^{(0)} + \gamma_{i}) \tilde{\Phi}^{(i)}(z,z_l)=0,\, z \ge z_l, \leqt{dPhi_i} \\
    \tilde{\Phi}^{(s)}(z_l,z_l) = - i \sqrt{2 \gamma_i} \Phi(z_l), \quad
    \tilde{\Phi}^{(i)}(z_l,z_l) = - i \sqrt{2 \gamma_s} \Phi(z_l) \leqt{dPhi_zl} ,
\end{eqnarray}
where $\Delta\beta^{(0)} = \beta_s^{(0)} + \beta_i^{(0)}$, and we take into account possible pump absorption with the loss coefficient $\gamma_p$ by putting $E_p(z) = A \exp(-\gamma_p z)$.
We disregard the evolution of $\tilde{\Phi}^{(si)}$ wavefunction, since it corresponds to the case when both photons are lost.

Equation~\reqt{dPhi} can be solved analytically:
\begin{eqnarray}
    \Phi(z)=&&z A {\rm sinc} \Bigg\{\frac{z}{z}\left[\Delta\beta^{(0)}-i(\gamma_{s}+\gamma_{i}-\gamma_{p})\right] \Bigg\}\nonumber\\
    &&\times \exp{\Bigg\{-\frac{iz}{2}\left[\Delta\beta^{(0)}-i(\gamma_{s}+\gamma_{i}+\gamma_{p})\right]\Bigg\}}. \leqt{Phi_slv}
\end{eqnarray}

We now calculate the normalized intensity of photons generated through SPDC, which is proportional to an average number of photons per unit time. The expressions for the signal and idler photons are analogous, and to be specific we consider the signal mode. The total signal intensity $I_s(z)$ is found as:
\begin{eqnarray}
    I_{s}(z)=I^{\left(0\right)}_{s}(z)+\tilde{I}_{s}(z),~
    I^{\left(0\right)}_{s}(z)=|\Phi(z)|^{2},~
    \tilde{I}_{s}(z)=\int^{z}_{0}dz_{l}\Bigg|\tilde{\Phi}^{(s)}(z,z_{l})\Bigg|^{2}, \leqt{Ints}
\end{eqnarray}
where $I^{\left(0\right)}_{s}(z)$ is the contribution when both photons are not absorbed and $\tilde{I}_{s}(z)$ is a contribution from the states with lost idler photons. Note that there is no interference between the photons with lost pairs~\cite{Zou:1991-318:PRL, Kulik:2004-31:JETP}.
The intensity contributions can be calculated analytically:
\begin{eqnarray}
    I^{\left(0\right)}_{s}=\frac{2A^{2}e^{-(\gamma_{s}+\gamma_{i}+\gamma_{p}) z}
    \           \left\{{\rm cosh}\left[(\gamma_{s}+\gamma_{i}-\gamma_{p}) z\right]-\cos\left(\Delta\beta^{(0)}z\right)\right\}}{
       \left(\Delta\beta^{(0)}\right)^2+(\gamma_{s}+\gamma_{i}-\gamma_{p})^{2}},\\
    \tilde{I}_{s}=\frac{4A^{2}\gamma_{i}e^{-2\gamma_{s}z}}{(\Delta\beta^{\left(0\right)})^2+(\gamma_{s}+\gamma_{i}-\gamma_{p})^{2}}
    \Big\{G\left[z,i(\gamma_{s}+\gamma_{i}-\gamma_{p})\right]-G(z,\Delta\beta^{(0)})\Big\},
\end{eqnarray}
where
\begin{eqnarray}
    G(z,p)=\int^{L}_{0}\cos(\xi p)e^{-\xi\left(\gamma_{i}+\gamma_{p}-\gamma_{s}\right)}d \xi \nonumber\\
    \quad = \frac{\gamma_{i}+\gamma_{p}-\gamma_{s}+e^{-z\left(\gamma_{i}+\gamma_{p}-\gamma_{s}\right)}
    \Bigg[p\sin(z p)-\cos(zp)(\gamma_{i}+\gamma_{p}-\gamma_{s})\Bigg]}{p^{2}+(\gamma_{i}+\gamma_{p}-\gamma_{s})^{2}}.
\end{eqnarray}
The total intensity can be measured by a sensitive camera, which will provide an overall number of detected photons per unit time.
The intensity contributions can be separated using a scheme with single-photon detectors: $I^{\left(0\right)}_{s}$ will be proportional to the number of coincidence counts of signal and idler photons, and $\tilde{I}_{s}$ will be proportional to the signal counts without the corresponding idler photon.

It is instructive to consider a number of limiting cases. In particular, zero pump loss ($\gamma_p = 0$) can be achieved in various conventional waveguides, where losses at pump frequency can be significantly smaller than losses at signal and idler frequencies due to the difference in the fundamental mode cross-section sizes for different wavelengths. In this case both components of signal intensity $I^{\left(0\right)}_{s}(z)$ and $\tilde{I}_{s}(z)$ approach stationary values for large distances:
\begin{eqnarray}\leqt{limitt}
    \lim_{z \rightarrow \infty} [I^{\left(0\right)}_{s}(z)]
    = \lim_{z \rightarrow \infty} [\tilde{I}_{s}(z)] \gamma_s \gamma_i^{-1}
    =\frac{A^2}{\left(\Delta \beta^{(0)}\right)^2 + (\gamma_s+\gamma_i)^2},
\end{eqnarray}
We see that if there is no idler loss ($\gamma_i=0$), then $\tilde{I}_{s}(z) \rightarrow 0$, which means that all signal photons are paired with an idler photon, as expected. If the signal and idler exhibit the same loss ($\gamma_s=\gamma_i$), then half of signal photons remains paired.

For degenerate SPDC regime with indistinguishable signal and idler photons ($\gamma_s=\gamma_i=\gamma$) and no pump losses ($\gamma_p = 0$), we have:
\begin{eqnarray}
    I^{\left(0\right)}_{s}(z)
    =\frac{2A^{2}e^{-2z\gamma}\left[{\rm cosh}(2z\gamma)-\cos(z\Delta\beta^{(0)})\right]}{(\Delta\beta^{\left(0\right)})^2+4\gamma^{2}},\\
    \tilde{I}_{s}(z) =\frac{2A^{2}e^{-2z\gamma}}{(\Delta\beta^{\left(0\right)})^2+4\gamma^{2}}\left[{\rm sinh}(2z\gamma)-2z\gamma{\rm sinc}(z\Delta\beta^{(0)})\right].
\end{eqnarray}

In the case of strongly non-degenerate SPDC, when signal and idler photons are generated with significantly different frequencies, pump and signal losses may become negligible $\gamma_p=\gamma_s=0$, while idler absorption may be substantial~\cite{Klyshko:1988:PhotonsNonlinear}. In this case the biphoton-related component of the signal intensity for long propagation distances is:
\begin{eqnarray} \leqt{aprInts}
    \lim_{z\rightarrow \infty}[ I^{\left(0\right)}_{s}(z)] = \frac{2 A_{s,i}\gamma_{i}}{(\Delta\beta^{\left(0\right)})^2+\gamma^{2}_{i}}.                                                                              \end{eqnarray}
We check that Eq.~\reqt{aprInts} agrees with the result derived in Ref.~\cite{Klyshko:1988:PhotonsNonlinear} through the application of fluctuation-dissipation theorem.

\pict{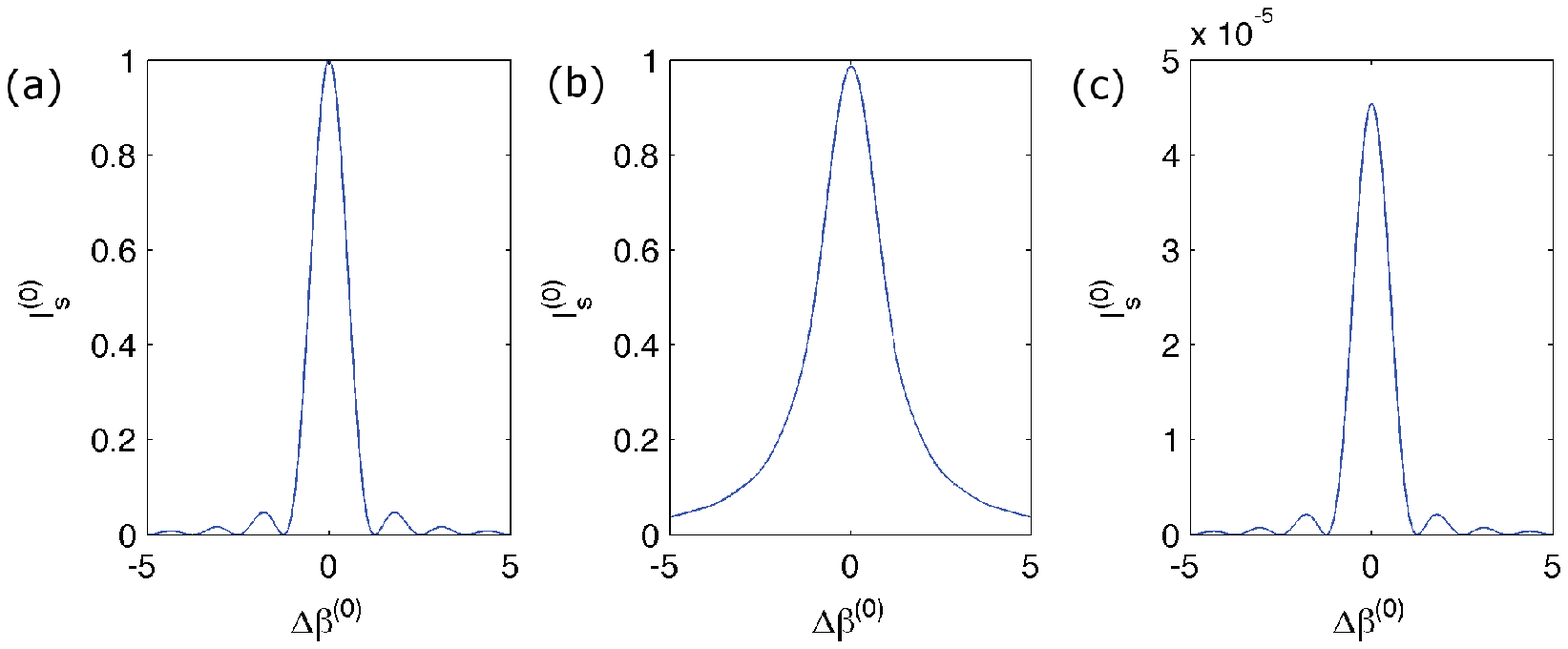}{sinc_lorenz}{
Normalized number of photon pairs, $I^{\left(0\right)}_{s}$, generated through SPDC in a single waveguide vs. the phase mismatch $\Delta\beta^{(0)}$ for $z=5$, $A=1$ and different losses: (a)~$\gamma_p = \gamma_s = \gamma_i = 0$, (b)~$\gamma_p = 0$, $\gamma_s = \gamma_i = 0.5$, (c)~$\gamma_s = \gamma_i = 0.5$, $\gamma_p = \gamma_s + \gamma_i=1$.
}

It is interesting to analyze the dependence of the biphoton-related component of the signal intensity $I^{\left(0\right)}_{s}$ on the phase mismatch $\Delta\beta^{(0)}$. When losses are absent ($\gamma_p = \gamma_s = \gamma_i = 0$), it has a well-known~\cite{Klyshko:1988:PhotonsNonlinear} shape of ${\rm sinc}$-function [Fig.~\rpict{sinc_lorenz}(a)]:
\begin{equation}
    I^{\left(0\right)}_{s}(z)=A^2 L^2 {\rm sinc}^2\left(\frac{\Delta \beta^{(0) } z}{2}\right) .
\end{equation}
For negligible pump losses ($\gamma_p = 0$) and large signal or idler losses \{$\exp{[-(\gamma_s+\gamma_i)z]} \ll 1$\} the dependence is transformed into a Lorenz shape [Fig.~\rpict{sinc_lorenz}(b)] according to Eq.~\reqt{limitt}.
Interestingly, when pump losses are increased to match the combined idler and signal losses ($\gamma_p = \gamma_s+\gamma_i$) the spectrum returns to a ${\rm sinc}$ shape [Fig.~\rpict{sinc_lorenz}(c)]:
\begin{equation}
    I^{\left(0\right)}_{s}(z)=A^2 z^2 \e^{-2 (\gamma_s+\gamma_i) z} {\rm sinc}^2\left(\frac{\Delta \beta^{(0) } z}{2}\right).
\end{equation}

\pict{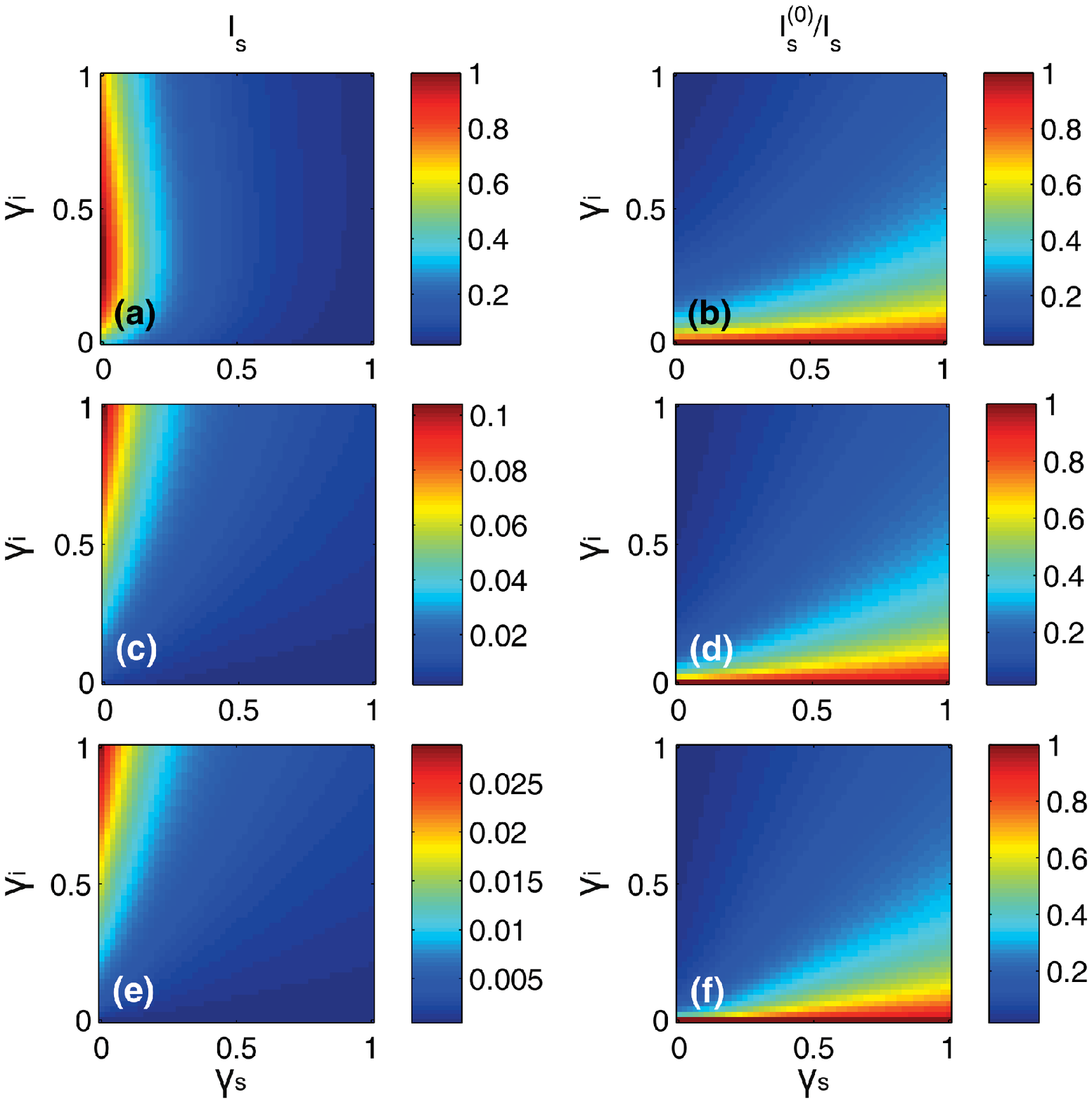}{1wgB}{
(a,c,e)~ total signal intensity $I_{s}(z)$ and (b,d,f)~ratio of intensity contribution when both photons are not absorbed and the full intensity $I^{\left(0\right)}_{s}(z) /I_{s}(z)$ vs. the signal and idler losses in a single waveguide for different values of phase mismatch (a,b)~$\Delta\beta^{(0)}=0$, (c,d)~$\Delta\beta^{(0)}=3$, (e,f)~$\Delta\beta^{(0)}=6$.
Parameters are $\gamma_p=0$, $z=5$, $A=1$.
}

\pict{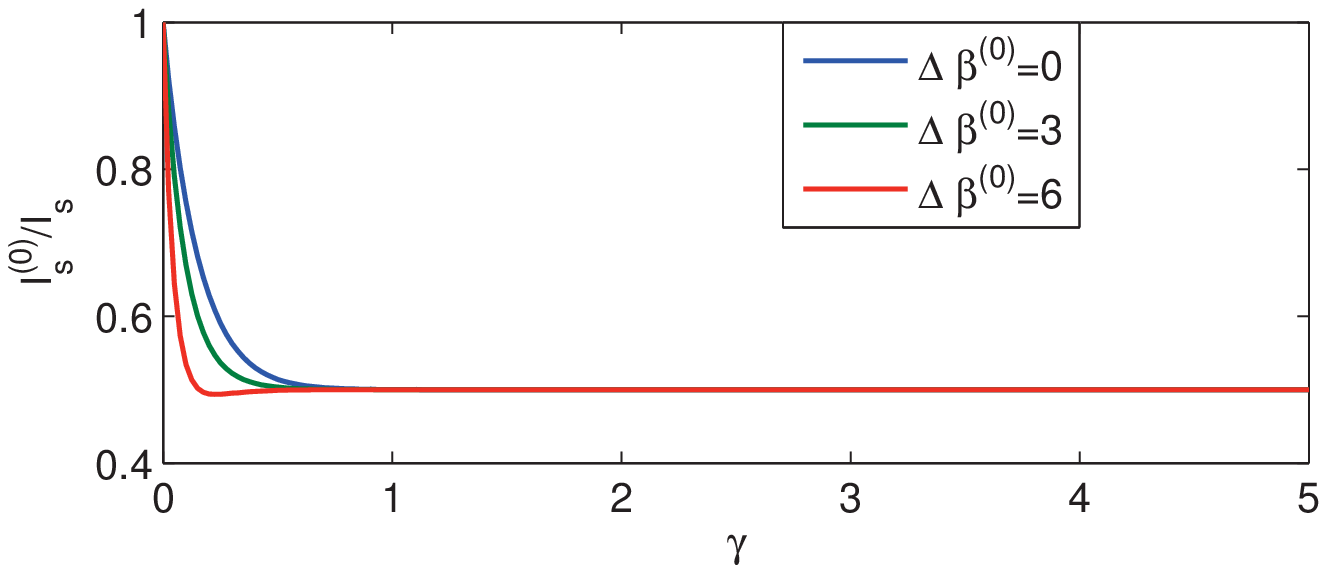}{1wgIg}{
Ratio of intensity contribution when both photons are not absorbed and the full intensity $I^{\left(0\right)}_{s}(z) / I_{s}(z)$ vs. the signal and idler loss $\gamma$ ($\gamma_{s}=\gamma_{i}=\gamma$, $\gamma_p=0$) in a single waveguide. Parameters are $A=1$,  $z=5$, and $\Delta\beta^{(0)}=\{0,3,6\}$ as indicated by labels.
}

Next we present a detailed investigation of the signal mode intensity depending on the loss (Figs.~\rpict{1wgB} and \rpict{1wgIg}) and propagation distance (Fig.~\rpict{1wgIz}) in the absence of pump loss $\gamma_p=0$. Figures~\rpict{1wgB}~(a,c,e) show that the signal intensity $I_s$ is decreasing with the increase of signal loss, however the dependence on the idler loss in relation to the phase mismatch $\Delta \beta^{(0)}$ is nontrivial due to additional signal intensity component $\tilde{I}_{s}$ related to the disruption of interference when the idler photon is lost. The ratio between the pure biphoton and the full signal intensity, $I^{\left(0\right)}_{s} / I_{s}$, depends weakly on the phase mismatch, see Figs.~\rpict{1wgB}~(b,d,f). Indeed, Fig.~\rpict{1wgIg} demonstrates that regardless of the phase mismatch $\Delta \beta^{(0)}$ the proportion of signal photons paired with idler to all signal photons, $I^{\left(0\right)}_{s} / I_{s}$, becomes independent on the loss above certain loss threshold.

Figure~\rpict{1wgIz} shows the behavior of the total signal intensity vs. the propagation distance for different losses in the regime of phase-matching. Total signal intensity exhibits fast growth in the absence of losses. However when moderate of high losses are present, the total signal intensity $I_s$ approaches a fixed value at large distances, see Eq.~\reqt{limitt}.

\pict{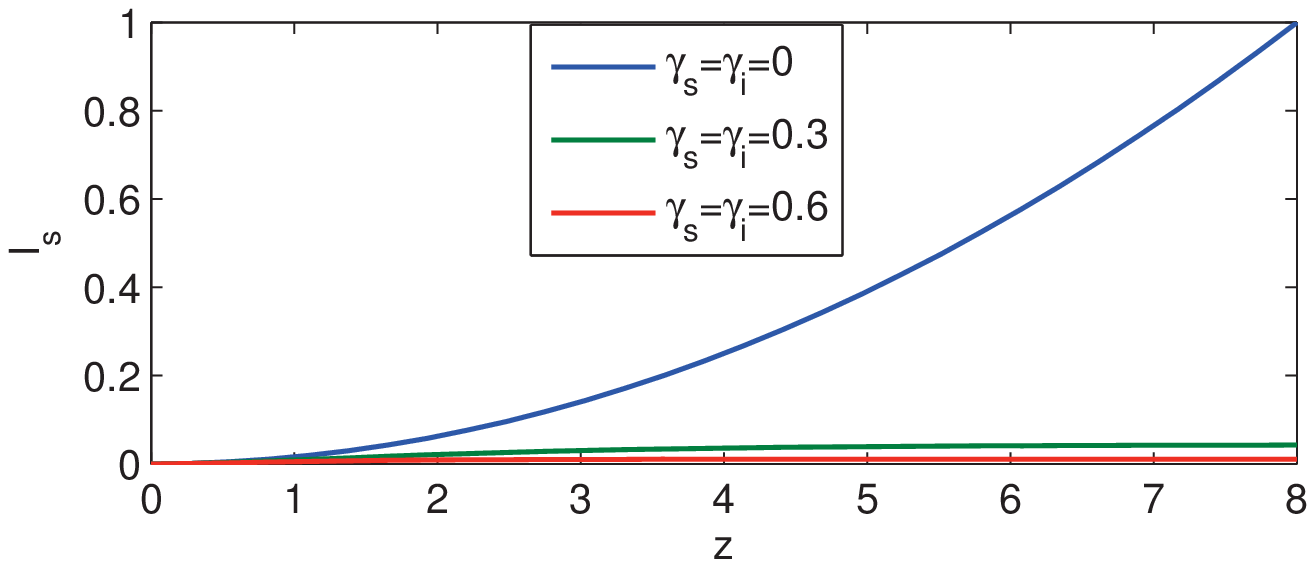}{1wgIz}{
Total signal mode intensity $I_s$ vs. the propagation distance in a single waveguide for different signal and idler losses $\gamma_{s}=\gamma_{i}=\gamma=\{0,0.3,0.6\}$. Parameters are $A=1$, $\Delta\beta^{(0)}=0$, $\gamma_p=0$.
}

\section{SPDC in Waveguide Array with Losses}\lsect{WGAS}

It was shown that nonlinear WGAs can serve as a reconfigurable on-chip source of spatially entangled photon pairs~\cite{Solntsev:2012-23601:PRL, Solntsev:2012-27441:OE, Grafe:2012-562:SRP, Kruse:2013-83046:NJP}.
Since internal generation of photon pairs in nonlinear waveguide arrays solves the problem of input losses, it is important to understand the effect of internal losses on photon-pair propagation and resulting entanglement and correlations.

For the theoretical analysis, we combine the one-waveguide Hamiltonians introduced in the previous section, and the linear coupling between the waveguides through the Hamiltonian $\hat{H}_{c}$. If the waveguide parameters are identical across the whole array, then the Hamiltonian is:
\begin{eqnarray}
  \leqt{Harray}
  \hat{H}(z) &=& \hat{H}_{nl}(z) + \hat{H}_{bs}(z) + \hat{H}_{c}(z), \\
  \hat{H}_{nl}(z) &=& \sum_{n_s} \beta_s^{(0)} a_s^\dag(n_s) a_s(n_s)
            +  \sum_{n_i} \beta_i^{(0)} a_i^\dag(n_i) a_i(n_i) \\
        &+& \sum_{n_p} \left[ E_p(z, n_p) a_s^\dag(n_p) a_i^\dag(n_p)
                                                  + E_p^\ast(z, n_p) a_s(n_p) a_i(n_p) \right] \nonumber \\
  \hat{H}_{bs}(z) &=& \sum_{n_s} \sqrt{2 \gamma_s} \left[ a_s(n_s) b_s^\dag(z, n_s)
                                                + a_s^\dag(n_s) b_s(z, n_s) \right]  \\
                      &+&   \sum_{n_i} \sqrt{2 \gamma_i} \left[ a_i(n_i) b_i^\dag(z, n_i)
                                    + a_i^\dag(n_i) b_i(z, n_i) \right]  \nonumber \\
  \hat{H}_{c}(z) &=& \sum_{n_s} C_s \left[ a_s(n_s) a_s^\dag(n_s+1)
                                                + a_s^\dag(n_s) a_s(n_s+1) \right]  \\
                   &+& \sum_{n_i} C_i \left[ a_i(n_s) a_i^\dag(n_i+1)
                                                + a_i^\dag(n_i) a_i(n_i+1) \right]  .
\end{eqnarray}
Here $n_s$ and $n_i$ are the waveguide numbers for the signal and idler photons, $a_{s,i}^\dag(n)$ and $a_{s,i}(n)$ are the creation and annihilation operators for the signal and idler photons in a waveguide number $n$, $b_{s,i}^\dag(z, n)$ describe creation of photons which are lost from a waveguide number $n$ at a coordinate $z$, $C_{s,i}$ are the coupling constants between the neighboring waveguides, $E_p(z, n_p)$
is proportional to pump amplitude in waveguide $n_p$.
Then, we seek a solution for a biphoton state as:
\begin{eqnarray}
        |\Psi(z)\rangle &=&
            \sum_{n_s} \sum_{n_i} \Phi_{n_{s},n_{i}}(z) a_s^\dag(n_s) a_i^\dag(n_i) |0\rangle \nonumber\\
         &+& \sum_{n_s} \sum_{n_i} \int_0^z d z_l \tilde{\Phi}_{n_{s},n_{i}}^{(s)}(z, z_l) a_s^\dag(n_s) b_i^\dag(z_l, n_i)  |0\rangle \nonumber\\
        & + & \sum_{n_s} \sum_{n_i} \int_0^z d z_l \tilde{\Phi}_{n_{s},n_{i}}^{(i)}(z, z_l) b_s^\dag(z_l, n_s) a_i^\dag(n_i)  |0\rangle \leqt{stateWGA} \\
        & + & \sum_{n_s} \sum_{n_i} \int_0^z d z_{l_s} \int_0^z d z_{l_i} \tilde{\Phi}_{n_{s},n_{i}}^{(si)}(z_{l_s}, z_{l_i}) b_s^\dag(z_{l_s}, n_s) b_i^\dag(z_{l_i}, n_i)  |0\rangle . \nonumber
\end{eqnarray}
The resulting set of equations for the evolution of the biphoton wave functions is:
\begin{eqnarray}
    \frac{\partial\Phi_{n_{s},n_{i}}(z)}{\partial z}
        &=&-i\Delta\beta^{(0)}\Phi_{n_{s},n_{i}}
        -(\gamma_{s}+\gamma_{i})\Phi_{n_{s},n_{i}}  + A_{n_{s}} \delta_{n_{s},n_{i}} e^{-\gamma_{p} z}\nonumber\\
         &-&i C_s(\Phi_{n_{s}-1,n_{i}}+\Phi_{n_{s}+1,n_{i}}) - i C_i (\Phi_{n_{s},n_{i}-1}+\Phi_{n_{s},n_{i}+1}) ,\\
    \frac{\partial\tilde{\Phi}_{n_{s},n_{i}}^{(s)}(z, z_l)}{\partial z}
        &=& - (i \beta_s^{(0)} + \gamma_{s}) \tilde{\Phi}_{n_{s},n_{i}}^{(s)}
        -i C_s (\tilde{\Phi}_{n_{s}-1,n_{i}}^{(s)}+\tilde{\Phi}_{n_{s}+1,n_{i}}^{(s)}), z \ge z_l,\\
    \frac{\partial\tilde{\Phi}_{n_{s},n_{i}}^{(i)}(z, z_l)}{\partial z}
        &=& - (i \beta_i^{(0)} + \gamma_{i}) \tilde{\Phi}_{n_{s},n_{i}}^{(i)}
        -i C_i (\tilde{\Phi}_{n_{s},n_{i}-1}^{(i)}+\tilde{\Phi}_{n_{s},n_{i}+1}^{(i)}), z \ge z_l,\\
    \tilde{\Phi}_{n_{s},n_{i}}^{(s)}(z_l, z_l) &=& - i \sqrt{2 \gamma_i} \Phi_{n_{s},n_{i}}(z_l), \,
    \tilde{\Phi}_{n_{s},n_{i}}^{(i)}(z_l, z_l) = - i \sqrt{2 \gamma_s} \Phi_{n_{s},n_{i}}(z_l) ,
\end{eqnarray}
where we do not consider the evolution of $\tilde{\Phi}^{(si)}$ wavefunction corresponding to both lost photons.
The real-space representation can be Fourier-transformed into spatial $k$-space~\cite{Grafe:2012-562:SRP}:
\begin{equation}
    \Phi_{k_{s},k_{i}}=\sum_{n_{s},n_{i}}\Phi_{n_{s},n_{i}} \e^{i n_{s} k_{s}} \e^{i n_{i} k_{i}}.
\end{equation}
Then the biphoton propagation equations in $k$-space can be written as follows:
\begin{eqnarray} \leqt{dPhi_k}
    \frac{\partial\Phi_{k_{s},k_{i}}}{\partial z}
    = -(i \Delta\beta + \gamma_{s}+\gamma_{i}) \Phi_{k_{s},k_{i}} + A_{k_{s},k_{i}}\e^{-\gamma_{p}z}, \\
    \frac{\partial\tilde{\Phi}_{n_{s},n_{i}}^{(s)}(z, z_l)}{\partial z}
    = - (i \beta_{s} + \gamma_s) \tilde{\Phi}_{n_{s},n_{i}}^{(s)},  \,
    \frac{\partial\tilde{\Phi}_{n_{s},n_{i}}^{(i)}(z, z_l)}{\partial z}
    = - (i \beta_{i} + \gamma_i) \tilde{\Phi}_{n_{s},n_{i}}^{(s)}, \leqt{dPhi_k_si}\\
    \tilde{\Phi}_{k_{s},k_{i}}^{(s)}(z_l, z_l) = - i \sqrt{2 \gamma_i} \Phi_{k_{s},k_{i}}(z_l), \,
    \tilde{\Phi}_{k_{s},k_{i}}^{(i)}(z_l, z_l) = - i \sqrt{2 \gamma_s} \Phi_{k_{s},k_{i}}(z_l) ,
\end{eqnarray}
where $\beta_s = \beta_s^{(0)} + 2 C_s \cos(k_s)$, $\beta_i = \beta_i^{(0)} + 2 C_i \cos(k_i)$, and $\Delta\beta = \beta_s + \beta_i$.
These equations have the same form as Eqs.~\reqt{dPhi}-\reqt{dPhi_zl} for a single waveguide. Accordingly, a solution for the wave function $\Phi_{k_{s},k_{i}}$ can be formulated analogous to Eq.~\reqt{Phi_slv}.

Finally, the real-space wave functions can be calculated by applying the inverse Fourier transform:
\begin{equation} \leqt{FTPs}
    \Phi_{n_{s},n_{i}}
    = \frac{1}{(2\pi)^2} \int^{\pi}_{-\pi}\int^{\pi}_{-\pi} {\rm d}k_{s}{\rm d}k_{i} \Phi_{k_{s},k_{i}}
                e^{-i k_{s}n_{s}} e^{-k_{i}n_{i}},
\end{equation}

We can also determine the reduced density matrixes, for instance, of the subsystem corresponding to the signal photons when idler photon is not lost, $\rho^{\left(0\right)}(k_{s1},k_{s2},z)$, and when the  idler photon is lost, $\widetilde{\rho_{s}}(k_{s1},k_{s2},z)$, as follows:
\begin{equation} \leqt{DMs}
    \rho^{\left(0\right)}(k_{s1},k_{s2},z)=\int {\rm d}k_{i}\Phi^{*}_{k_{s1},k_{i}}(z)\Phi_{k_{s2},k_{i}}(z),
\end{equation}
\begin{equation} \leqt{DMts}
    \widetilde{\rho_{s}}(k_{s1},k_{s2},z)=\int^{z}_{0}{\rm d}z_{l}\int {\rm d}k_{i}\left[\tilde{\Phi}^{\left(s\right)}_{k_{s1},k_{i}}(z,z_{l})\right]^{*}\tilde{\Phi}^{\left(s\right)}_{k_{s2},k_{i}}(z,z_{l}).
\end{equation}
Taking into account Eq.~\reqt{dPhi_k} and Eq.~\reqt{dPhi_k_si} we can write the master equations for $\widetilde{\rho_{s}}(k_{s1},k_{s2},z)$ as:
\begin{equation} \leqt{RoS}
    \frac{\partial\widetilde{\rho_{s}}(k_{s1},k_{s2},z)}{\partial z} = 2\gamma_{i}\rho^{\left(0\right)}(k_{s1},k_{s2},z)-2\gamma_{s}\widetilde{\rho_{s}}(k_{s1},k_{s2},z).
\end{equation}
This equation represents the propagation of the signal photon with the lost idler photon, where the first term corresponds to the probability of the idler photon loss, while the second term accounts for the possibility of the signal photon to be lost as well.
\pict{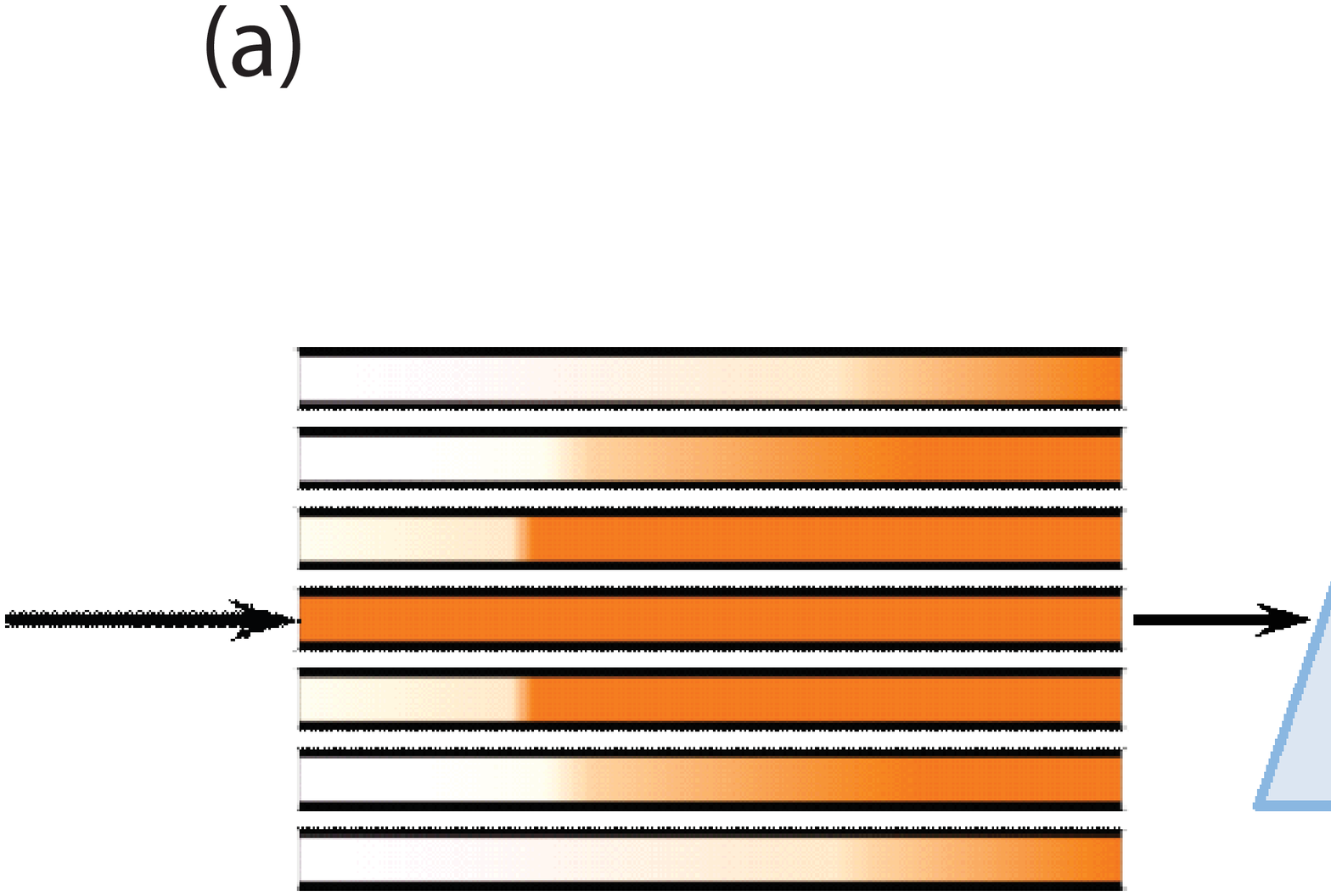}{MWGA}{
Scheme of the experimental setup designed to measure spectral and spatial distribution of the nonlinear WGA output photon-pair intensity. The pump beam generates photon pairs that suffer losses and couple to the neighboring waveguides. (a)~ the output intensity distribution can be characterised using a prism and a camera. Spectral filtering can be used to choose only a signal channel to measure the signal intensity $I_s$. (b)~ the photon-pair correlations can be characterised by measuring the coincidences from the two single photon detectors.
}

The dependence of the intensity for the signal mode on the propagation distance can be written in the following form for $k$-space:
\begin{eqnarray}
    I_{s}(k_{s},z)=I^{\left(0\right)}_{s}(k_{s},z)+\tilde{I}_{s}(k_{s},z) ,
\end{eqnarray}
\begin{eqnarray} \leqt{IntsWGAk}
    I^{\left(0\right)}_{s}(k_{s},z)&=&\int^{\pi}_{-\pi}{\rm d}k_{i}|\Phi_{k_{s},k_{i}}(z)|^{2}, \\
    \tilde{I}_{s}(k_{s},z) & =& \int^{z}_{0}{\rm d}z_{l} \int^{\pi}_{-\pi}{\rm d}k_{i}\Bigg|\tilde{\Phi}^{(s)}_{k_{s},k_{i}}(z,z_{l})\Bigg|^{2} ,
\end{eqnarray}
and analogously for real space:
\begin{eqnarray}
    I_{n_{s}}(z)=I_{s}^{\left(0\right)}(n_{s},z) + \tilde{I}_{s}(n_{s},z),
\end{eqnarray}
\begin{eqnarray}
    I_{s}^{\left(0\right)}(n_{s},z)=\sum_{n_{i}}|\Phi_{n_{s},n_{i}}(z)|^{2},
    \tilde{I}_{s}(n_{s},z) = \int^{z}_{0}{\rm d}z_{l}\sum_{n_{i}}\Bigg|\tilde{\Phi}^{(s)}_{n_{s},n_{i}}(z,z_{l})\Bigg|^{2}.
\end{eqnarray}
Here $I_{s}^{\left(0\right)}(n_{s},z)$ is the contribution when both photons are not absorbed, and $\tilde{I}_{s}(n_{s},z)$ is a contribution from the states with lost idler photons.

In experiments, total intensity can be characterised for various wavelengths by using a spatially-resolving spectrometer [Fig.~\rpict{MWGA}(a)], which allows one to measure the signal and idler intensity outputs from different waveguides at various frequencies. Additionally, a coincidence scheme at the WGA output [Fig.~\rpict{MWGA}(b)] can be used to measure biphoton spatial correlations~\cite{Peruzzo:2010-1500:SCI, Solntsev:2012-23601:PRL}, which normalized value is:
\begin{equation}
    \Gamma_{n_{s},n_{i}}(z)=|\Phi_{n_{s},n_{i}}(z)|^{2}.
\end{equation}

\pictb{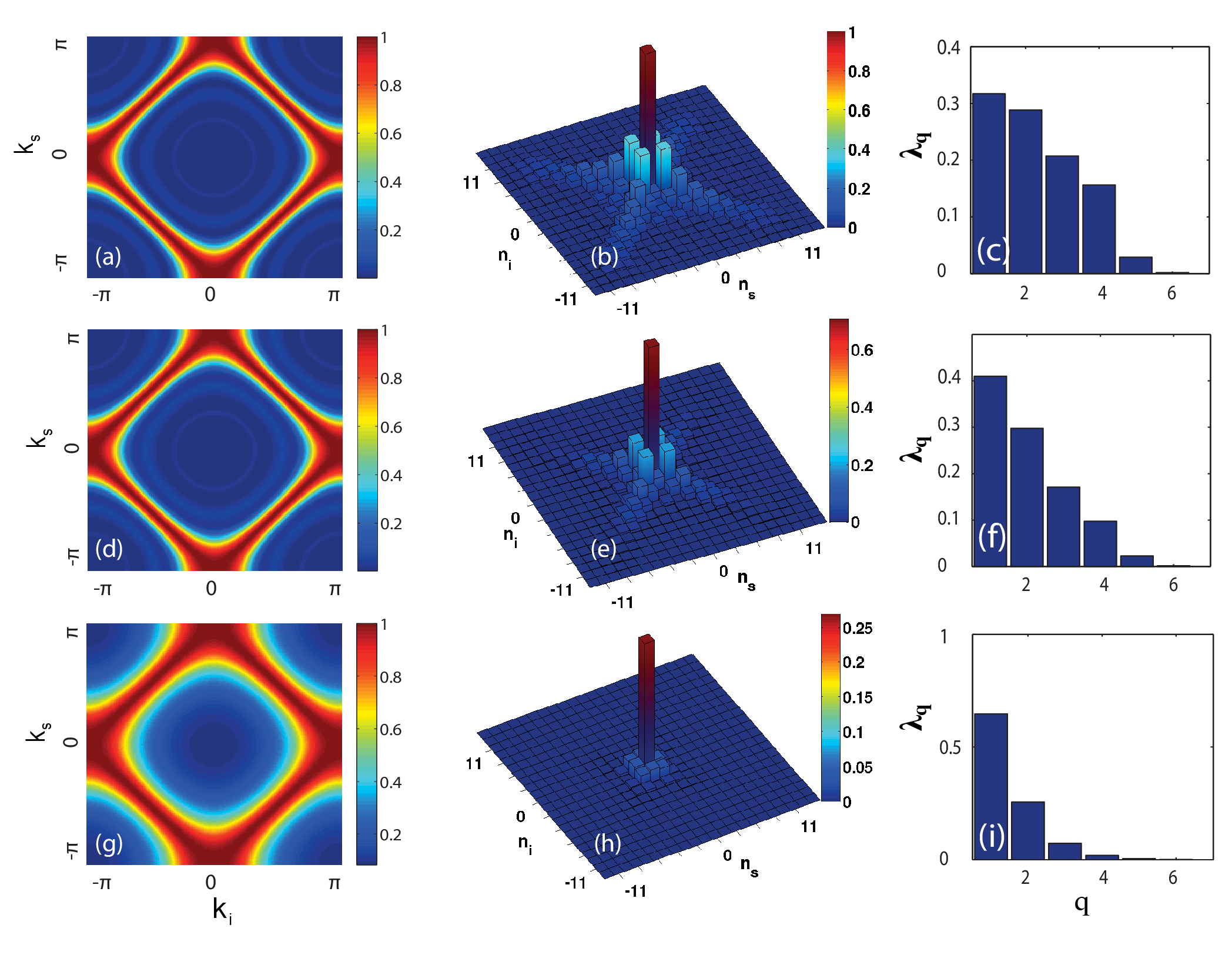}{N0Cor}{
Photon-pair correlations in (a,d,g)~$k$-space; (b,e,h)~real space correlations and (c,f,i)~ Schmidt decomposition depending on the mode number for different signal and idler loss, $\gamma_s=\gamma_i=\gamma/2$:
(a,b,c)~$\gamma=0$, (d,e,f)~$\gamma=0.2$, (g,h,i)~$\gamma=0.6$.
The pump is coupled to the central waveguide, $A(0)=0$. Parameters are $z=5$, $C_s=C_i=1$, $\gamma_p=0$, $\Delta\beta^{(0)}=0$.
}

We present the plots of photon-pair correlations in $k$-space and real space for different values of losses in Fig.~\rpict{N0Cor}, considering the pump beam coupled to a single waveguide. In $k$-space [Fig.~\rpict{N0Cor}(a,d,g)] at different loss values the correlation profiles have a square shape corresponding to angular phase-matching in waveguide arrays~\cite{Solntsev:2012-23601:PRL, Kruse:2013-83046:NJP}, however the square edges become broader for higher losses.
The real-space correlations in the absence of loss have a characteristic cross shape corresponding to the simultaneous bunching and anti-bunching  [Fig.~\rpict{N0Cor}(b)], which is a signature of non-classicality~\cite{Peruzzo:2010-1500:SCI, Solntsev:2012-23601:PRL}. Importantly, these non-classical features are preserved even in presence of moderate loss [Fig.~\rpict{N0Cor}(e)]. Under strong loss, photons are only present in the central waveguide [Fig.~\rpict{N0Cor}(h)], as photons are absorbed before they can couple to the neighboring waveguides, and accordingly the non-classical spatial correlations are absent.

To exploit two photons as quantum resources it is necessary to know if they are entangled.  This question can be answered by studying the Schmidt decomposition~\cite{Ekert:1995-415:AMJP} of a biphoton wave function
as follows:
\begin{equation}\leqt{ScD}
    \Phi_{k_{s},k_{i}}=\sum_{q} \sqrt{\lambda_{q}} \phi_{q}(k_{s}) \varphi_{q}(k_{i}),
\end{equation}
where $\lambda_{q}$ are Schmidt coefficients ($\sum_{q} \lambda_{q}=1$), and $\phi_{q}(k_{s})$ and $\varphi_{q}(k_{i})$ are Schmidt functions.

As mentioned previously, the generated photon pairs couple to a smaller number of neighboring waveguides with the increase of losses [Fig.~\rpict{N0Cor}(b,e,h)]. The same dynamics is also seen from the plots of Schmidt decomposition, where a single mode becomes dominating and the spatial entanglement decreases while losses increase [Fig.~\rpict{N0Cor}(c,f,i)].

The output photon statistics can be tailored by changing the pump profile and phase. When the pump beam is
coupled with equal amplitudes and phases to two neighboring waveguides, $A(n)=1$ for $n=0,1$, then the
correlations are strongly modified (Fig.~\rpict{PhCor}) compared to the single-waveguide pump excitation (Fig.~\rpict{N0Cor}). As losses increase, the photon-pair correlations are broadened in $k$-space [Fig.~\rpict{PhCor}(a,d,g)] and gradually fade in real space~[Fig.~\rpict{PhCor}(b,e,h)]. An interesting point here is that with the increase of losses the real-space correlations transform from predominantly antibunching pattern (with the largest correlations on the anti-diagonal, $n_{s}=-n_{i}$) at low and moderate loss [Fig.~\rpict{PhCor}(b,e)] to bunching pattern (with the largest correlations on the diagonal, $n_{s}=n_{i}$) at high loss [Fig.~\rpict{PhCor}(h)]. The Schmidt decomposition in the case of pump in two neighbouring waveguides [Fig.~\rpict{PhCor}(c,f,i)] shows the dynamics, which is similar to that in the case of pump in a single waveguide [Fig.~\rpict{N0Cor}(c,f,i)], although the Schmidt modes are distributed in pairs .

\pict{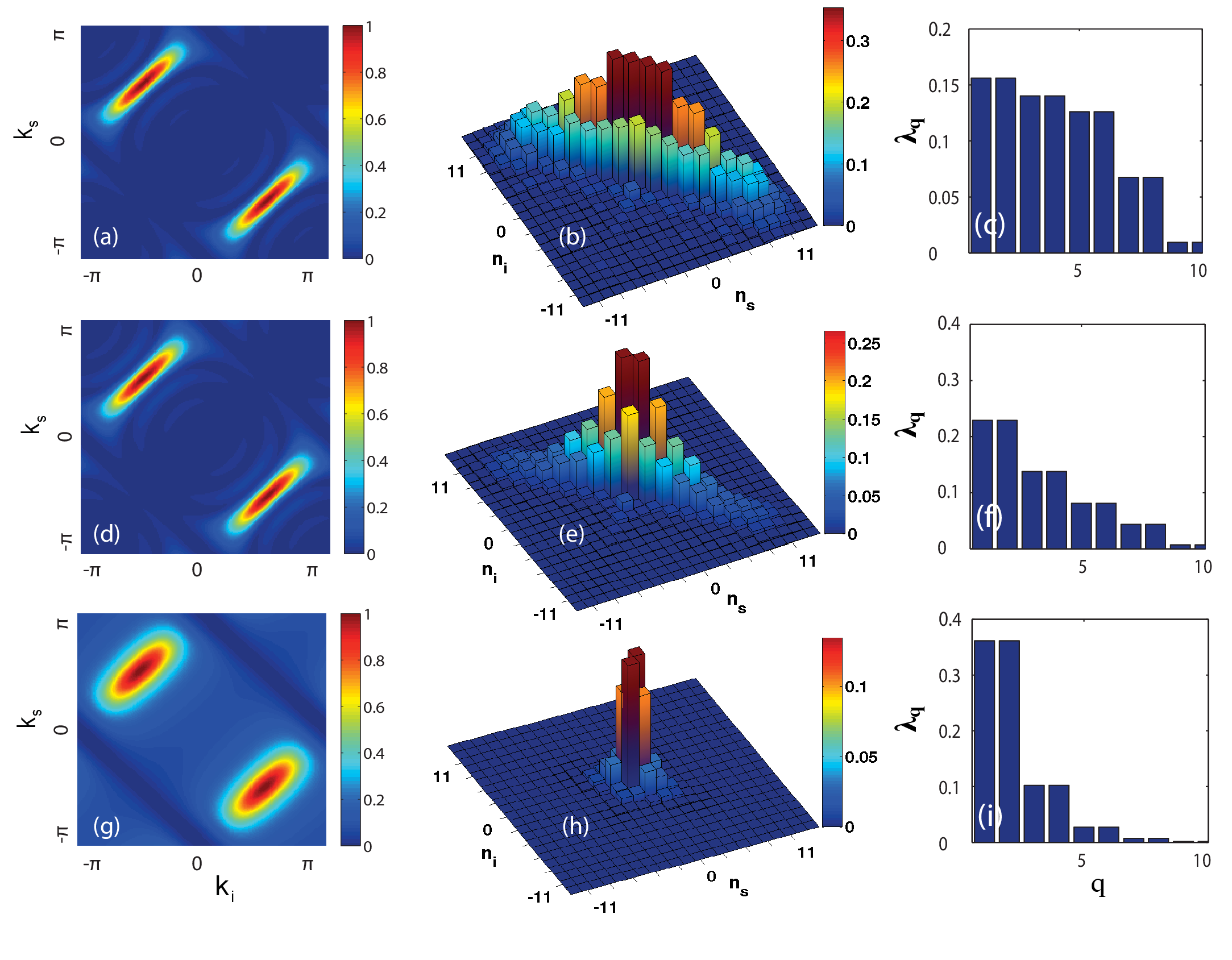}{PhCor}{
Photon-pair correlations in (a,d,g)~$k$-space; (b,e,h)~real space correlations and (c,f,i)~ Schmidt decomposition depending on the mode number (see, Eq.~\reqt{ScD}) for different signal and idler loss, $\gamma_s=\gamma_i=\gamma/2$: (a,b,c)~$\gamma=0$, (d,e,f)~$\gamma=0.2$, (g,h,i)~$\gamma=0.6$.
The pump is coupled in-phase to two neighboring waveguides in the centre, $A(0)=A(1)=1$. Parameters are $z=5$, $C_s=C_i=1$, $\gamma_p=0$, $\Delta\beta^{(0)}=0$.
}

\pict{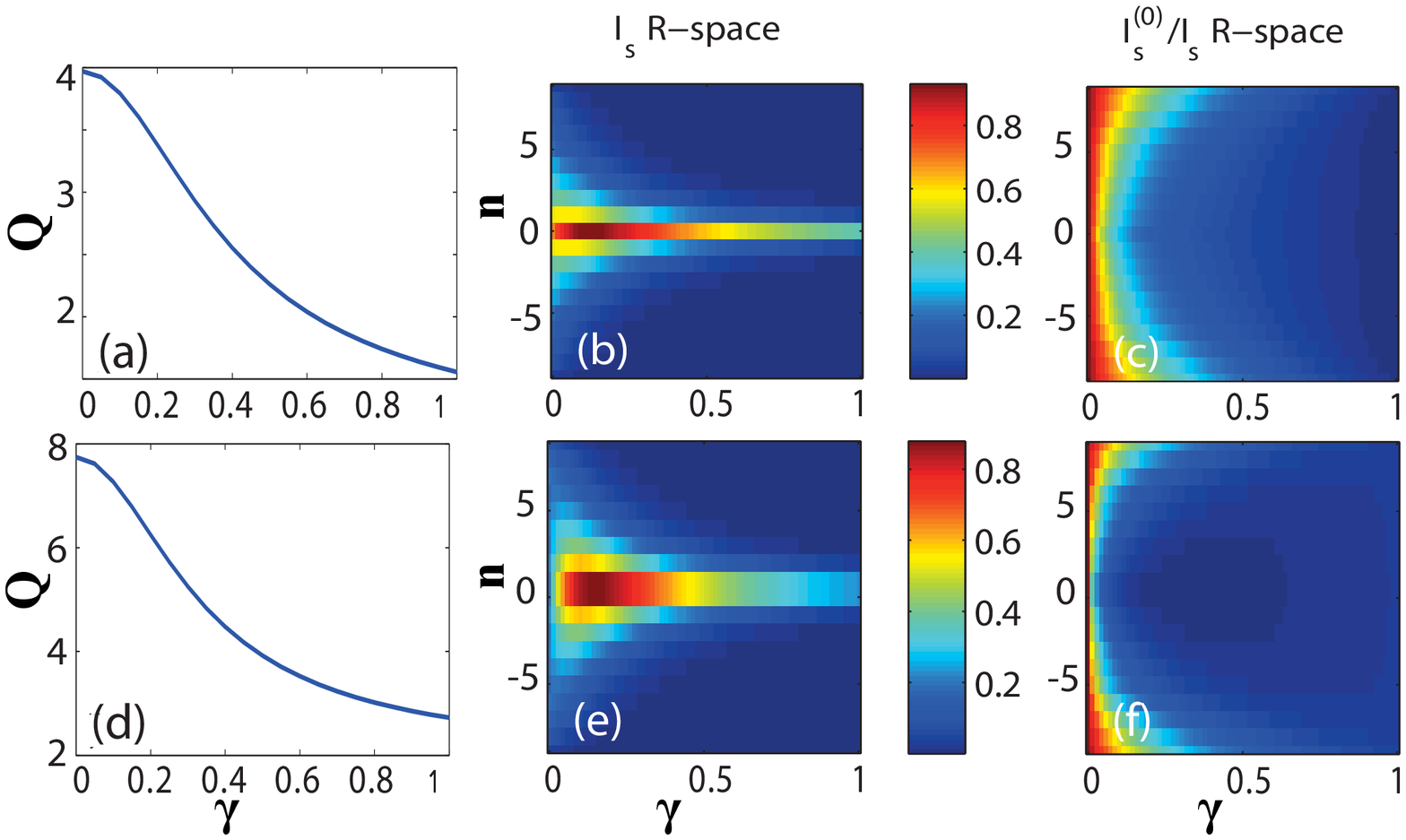}{Gam}{
(a,d)~Schmidt number, (b,e)~the signal mode full intensity in real space and (c,f)~fraction of signal photons coupled with idler photons to all signal photons in real space vs. the signal and idler loss $\gamma$ ($\gamma_{i}=\gamma_{s}=\gamma/2$) for different pump profiles: (a-c)~pump coupled to the central waveguide, $A(0)=1$, (d-f)~pump coupled in-phase to two neighboring waveguides, $A(0)=A(1)=1$. Parameters are $z=5$, $C_s=C_i=1$, $\gamma_p=0$, $\Delta\beta^{(0)}=0$.
}

\pict{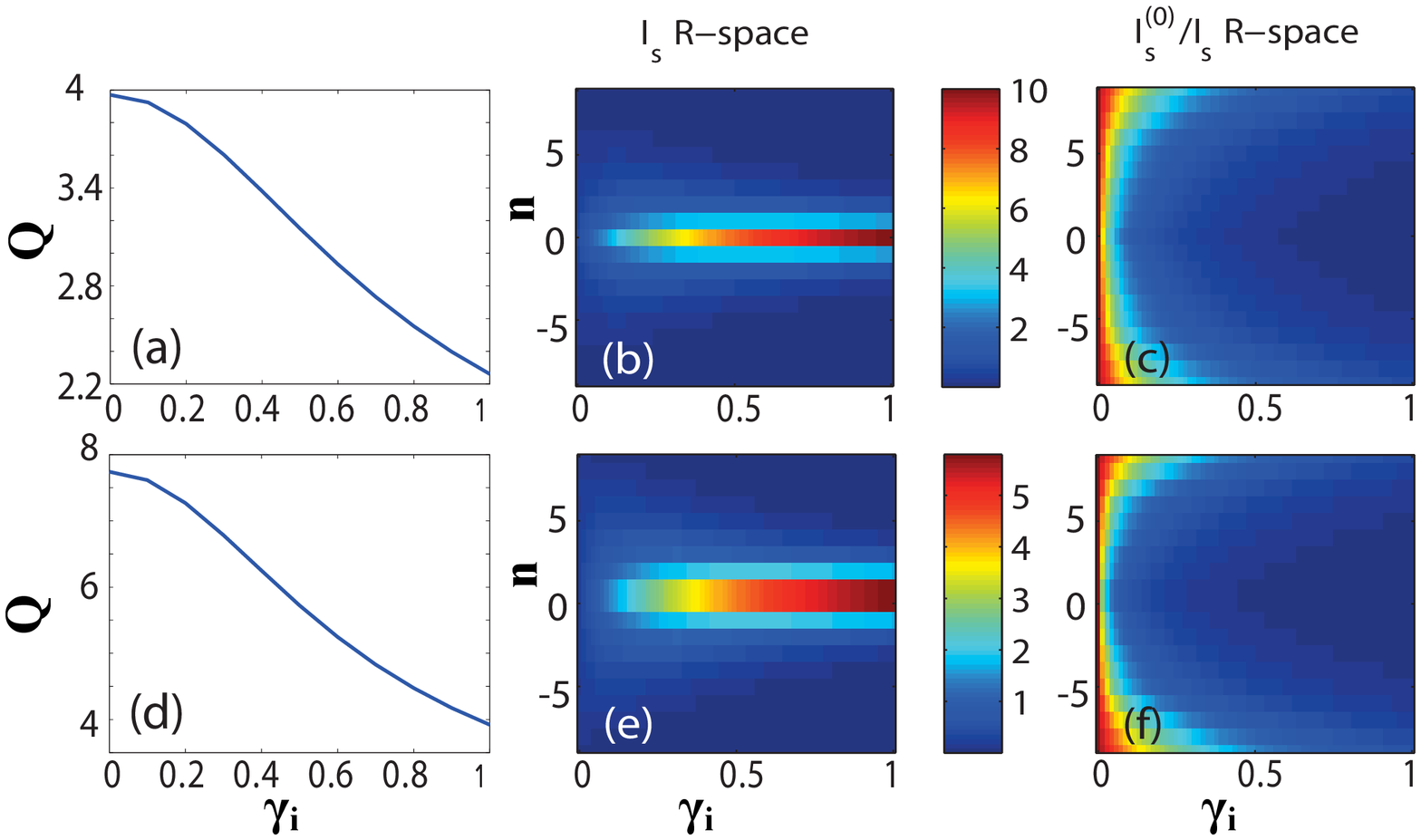}{GamI}{
(a,d)~Schmidt number, (b,e)~the signal mode full intensity in real space and (c,f)~fraction of signal photons coupled with idler photons to all signal photons in real space vs. the idler loss $\gamma_i$ for different pump profiles: (a-c)~pump coupled to the central waveguide, $A(0)=1$, (d-f)~pump coupled in-phase to two neighboring waveguides, $A(0)=A(1)=1$. Parameters are $z=5$, $C_s=C_i=1$, $\gamma_p=\gamma_s=0$, $\Delta\beta^{(0)}=0$.
}

The amount of entanglement can be conveniently quantified by the cooperativity parameter~-- Schmidt number $Q$~\cite{Huang:1993-915:JMO, URen:2005-146:LP}, which is defined in terms of Schmidt eigenvalues as follows:
\begin{equation}
    Q=\frac{1}{\sum_{q}\lambda^{2}_{q}}.
\end{equation}
The lowest value of $Q=1$ corresponds to a system with no quantum entanglement.

We show the dynamics of the Schmidt number [Fig.~\rpict{Gam}(a,d)] as well as full signal intensity in real space [Fig.~\rpict{Gam}(b,e)] and the ratio $I^{\left(0\right)}_{s}/I_{s}$ defining the fraction of signal photons coupled with the idler photon to all signal photons [Fig.~\rpict{Gam}(c,f)] for different values of loss $\gamma$ ($\gamma_{s}=\gamma_{i}=\gamma/2$) and two different pump excitations. We see that both in the case when pump is coupled to the central waveguide and in the case when pump is coupled to two neighboring waveguides, the total signal intensity $I_{s}$ first increases and then starts to decrease with the increase of losses, while the Schmidt number and the ratio $I^{\left(0\right)}_{s}/I_{s}$ always decrease with the increase of losses.

It is also interesting to consider the case of non-degenerate SPDC, when there is no signal loss ($\gamma_{s}=0$), and only the idler loss is present ($\gamma_{i} > 0$). We show the corresponding Schmidt number, the total signal intensity $I_{s}$ and signal intensity ratio $I^{\left(0\right)}_{s}/I_{s}$ in Fig.~\rpict{GamI}. In this case while the Schmidt number~[Fig.~\rpict{Gam}(a,d)] and the ratio $I^{\left(0\right)}_{s}/I_{s}$~[Fig.~\rpict{Gam}(c,f)] decrease, total signal intensity $I_{s}$~[Fig.~\rpict{Gam}(b,d)] always increases with the increase of idler losses. These trends are in agreement with the single waveguide case [c.f. Fig.~\rpict{1wgB}], however in a waveguide array we additionally observe loss-influenced reshaping of spatial intensity profiles.

\section{Conclusion}
In this work we have performed analytical and numerical analysis of the effect of linear losses on spontaneous parametric down-conversion in quadratic nonlinear waveguide and waveguide arrays, considering in detail biphoton and single-photon outputs under a variety of conditions. We have shown that idler losses can lead to increase of signal intensity and stabilisation of signal output in relation to the waveguide length. We have also demonstrated that signal and idler losses lead to the transformation of common sinc-shaped photon-pair correlation spectrum into a Lorenzian shape, and that this transformation can be fully reversed by specific increase in pump losses. Finally we have shown that nonlinear waveguide arrays can serve as a robust integrated platform for the generation of entangled photon states with non-classical spatial correlations, and that the operation of such quantum circuit is tolerant even to relatively high losses. We expect that this work will open new opportunities in developing loss-tolerant quantum integrated circuits.

\section{Ankowledgements}
The work has been supported by the Australian Research Council, including Discovery Project DP130100135 and Future Fellowship FT100100160.

\section*{References}

\providecommand{\newblock}{}

\end{document}